\documentstyle[twocolumn,prd,aps,epsf]{revtex}

\preprint{NUC-MINN-01/7-T}

\newcommand{\be}{\begin{equation}}
\newcommand{\ee}{\end{equation}}
\newcommand{\ba}{\begin{eqnarray}}
\newcommand{\ea}{\end{eqnarray}}

\begin{document}
\draft

\title{Carlson-Goldman modes in the color superconducting phase of
dense QCD}

\author{V. P. Gusynin$^{1}$ and I. A. Shovkovy$^{2,*}$}
\address{$^{1}$Bogolyubov Institute for Theoretical Physics,
        252143, Kiev, Ukraine\\
    and Department of Physics, Nagoya University,
        Nagoya 464-8602, Japan}

\address{$^{2}$School of Physics and Astronomy, University of Minnesota,
        Minneapolis, MN 55455, USA}

\maketitle

\begin{abstract}
We predict the existence of the gapless Carlson-Goldman modes in the
color-flavor locked color superconducting phase of dense QCD. These modes
resemble Nambu-Goldstone modes of the superconducting phase below the
critical temperature where the Anderson-Higgs mechanism takes place. These
modes exist in the broken phase in the vicinity of the critical line.
Their presence does not eliminate the Meissner effect. The effect of
Landau damping on the width of the Carlson-Goldman modes is discussed.
\end{abstract}




The Anderson-Higgs mechanism is one of the most fundamental notions of
modern theoretical physics, having many important applications in particle
and condensed matter physics. The Standard Model of particle physics,
which has already withstood numerous experimental tests, crucially
incorporates the Higgs mechanism as one of the most essential ingredients.
Similarly, the theory of the low and high temperature superconductivity
would not be complete without the Anderson mechanism.

About a three decades ago, an unusual and a rather surprising
``propagating order parameter collective modes" were discovered by Carlson
and Goldman \cite{CG} (see also, Ref.~\cite{Kulik,Tak1,Tak2}). One of the
most intriguing interpretation connects such modes with a revival of the
Nambu-Goldstone (NG) bosons in the superconducting phase where the
Anderson-Higgs mechanism should commonly take place. It was argued that an
interplay of two effects, screening and Landau damping, is crucial for the
existence of the CG modes \cite{Tak1,Tak2}. These modes can only
appear in the vicinity of the critical temperature (in the broken phase),
where a large number of thermally excited quasiparticles leads to partial
screening of the Coulomb interaction, and the Anderson-Higgs mechanism
becomes inefficient. At the same time, the quasiparticles induce Landau
damping which usually makes the CG modes overdamped in clean systems. To
suppress such an effect and make the CG modes observable, one should
consider dirty systems in which quasiparticle scatterings on impurities
tend to reduce Landau damping.

In the two fluid description, the CG modes are related to oscillations of
the superfluid and the normal component in opposite directions
\cite{SchSch}. The local charge density remains zero in such oscillations,
providing favorable conditions for gapless modes, in contrary to the
widespread belief that the plasmons are the only collective modes in
charged systems.

In this Letter we predict the existence of the Carlson-Goldman (CG) modes
in dense quark matter. It has been known for a long time that quark matter
at a sufficiently high baryon density should reveal color
superconductivity \cite{BarFra,Bail}. Recent developments \cite{W1,S1}
added a hope that a color superconducting phase may exist in compact
stars, and its signatures could possibly be detected \cite{HIC-Stars}.

We shall consider the color-flavor locked (CFL) phase of dense QCD with
three massless flavors of quarks (up, down and strange).  In this phase,
the original $SU(3)_{C} \times SU(3)_{R} \times SU(3)_{L} $ symmetry of
the QCD action breaks down to $SU(3)_{C+L+R}$ \cite{ARW}. Out of total
sixteen (would be) NG bosons, eight ($\phi^{A}$, $A=1,\ldots, 8$) are
removed from the physical spectrum by the Higgs mechanism, providing
masses to eight gluons. The other eight NG bosons ($\pi^{A}$) show up as
an octet of pseudoscalars. In addition, the global baryon number symmetry
as well as the approximate $U(1)_{A}$ symmetry also get broken. This gives
an extra NG boson and a pseudo-NG boson.

The order parameter consists of a color antitriplet, flavor antitriplet
$(\bar{3},\bar{3})$ and a color sextet, flavor sextet $(6,6)$
contributions \cite{PR1}. The dominant contribution to the order parameter
is $(\bar{3},\bar{3})$, but a non-zero, although small, admixture of
$(6,6)$ also appears \cite{ARW,us2,Spt}. The value of the order parameter
at zero temperature was estimated using phenomenological \cite{W1,S1}, as
well as microscopic models \cite{Son,us,SW2,PR-weak}.  While the exact
value remains uncertain, it can be as large as $100$ MeV.  Without loosing
generality, here we consider a pure $(\bar{3},\bar{3})$ order parameter,
so that the quark gaps in the octet and singlet channel are related as
follows $\Delta_{1}\approx -2\Delta_{8}=-2|\Delta_{T}|$ at any finite
temperature. As we shall see below, the presence of two distinct (octet
and singlet) quark gaps in the CFL phase leads to a partial suppression of
Landau damping, and, as a result, the CG modes could show up even in a
clean color superconductor.

Our current analysis mostly deals with the collective modes, related to
the gluon field. It is natural, then, to start our consideration with the
effective action, obtained by integrating out the quark degrees of
freedom. Irrespective of a specific model, the corresponding Lagrangian
density should read
\ba
{\cal L}&=& -\frac{1}{2} A^{A,\mu}_{-q}
i \left[{\cal D}^{(0)}(q)\right]^{-1}_{\mu\nu} A^{A,\nu}_{q}
\nonumber \\
&-&\frac{1}{2}\left[A^{A,\mu}_{-q} -iq^{\mu}\phi^{A}_{-q} \right]
\Pi_{\mu\nu}(q)
\left[A^{A,\nu}_{q}+iq^{\nu}\phi^{A}_{q} \right] +\ldots,
\label{L-g-phi}
\ea
where ellipsis denote the interactions terms.  The presence of the phase
field octet $\phi^{A}_{q}$ is very important for preserving the gauge
invariance of the model. Under a gauge transformation, the phase
$\phi^{A}_{q}$ in Eq.~(\ref{L-g-phi}) is shifted so that it would
exactly compensate the transformation of the gluon field. (Here we
consider only infinitesimally small gauge transformations. In general,
the gauge transformation of $\phi^{A}_{q}$ field is not a simple shift.
However, it is always true that its transformation exactly compensates
the transformation of the gluon field.)

In the simplest approximation, the polarization tensor $\Pi_{\mu\nu}$ in
Eq.~(\ref{L-g-phi}) is given by a one-loop quark diagram. Its formal
expression was presented in Ref.~\cite{Rsch1} (also, the zero temperature
limit was considered in Ref.~\cite{Zar}, and some other limits will be
presented in Ref.~\cite{GS-long}).  Let us discuss the general properties
of this tensor.

We start by pointing out that the one-loop polarization tensor
$\Pi_{\mu\nu}$ in Eq.~(\ref{L-g-phi}) obtained by integrating out the
quark degrees of freedom is {\em not} transverse in general. This is
directly related to the Higgs effect in dense QCD. The latter implies the
presence of the composites with the quantum numbers of the (would be) NG
bosons in all nonunitary gauges \cite{bs-23}. In the problem at
hand, in particular, this means that the complete expression for the
polarization tensor should necessarily contain an additional contribution
coming from integrating out the (would be) NG bosons [denoted by
$\phi^{A}_{q}$ fields in Eq.~(\ref{L-g-phi})]. Having said this, we should
note right away that the one-loop expression for $\Pi_{\mu\nu}$ {\em is}
transverse in the normal phase of the quark matter (above $T_{c}$) where
there are no (would be) NG bosons.

The longitudinal part of the one-loop polarization tensor $\Pi_{\mu\nu}$
has a physical meaning in the Higgs phase of dense QCD. It was shown in
Ref.~\cite{Zar} that such a longitudinal part contains the information
about the propagators of the pseudoscalar NG bosons, and, thus, determines
the corresponding dispersion relations. Strictly speaking, of course, the
properties of the NG bosons are related to the ``axial" polarization
tensor. In the model at hand, the two quantities are identical to the
leading order, and we could safely interchange them.

The action in Eq.~(\ref{L-g-phi}) is a starting point in our analysis.
In order to construct the current-current correlation function, we
should add the classical external gauge fields to the action in
Eq.~(\ref{L-g-phi}) and integrate out all quantum fields. After doing so,
we obtain the generating functional. The correlation function, then, is
given by the second order derivative with respect to the external gauge
field. In momentum space, the result reads \cite{GS-long}
\ba
\langle j^{A}_{\mu} j^{B}_{\nu} \rangle_{q}
&=& \delta^{AB} \left[
\frac{ q^{2} \Pi_{1}}{q^{2}+\Pi_{1}}
 O^{(1)}_{\mu\nu}(q) \right. \nonumber \\
&+& \left.
\frac{ q^{2} \left[\Pi_{2} \Pi_{3} + (\Pi_{4})^{2}\right]}
{(q^{2}+ \Pi_{2}) \Pi_{3} + (\Pi_{4})^{2}}
O^{(2)}_{\mu\nu}(q)
\right]. \label{cur-cur}
\ea
Here $\Pi_{i}$ ($i=1,\ldots 4$) are the component functions of the
one-loop polarization tensor, introduced as follows:
\be
\Pi_{\mu\nu}(q) =
\Pi_{1} O^{(1)}_{\mu\nu}
+\Pi_{2} O^{(2)}_{\mu\nu}
+\Pi_{3} O^{(3)}_{\mu\nu}
+\Pi_{4} O^{(4)}_{\mu\lambda},
\label{Pi-gen}
\ee
and
\ba
O^{(1)}_{\mu\nu} (q) &=& g_{\mu\nu}-u_{\mu} u_{\nu}
+\frac{\vec{q}_{\mu}\vec{q}_{\nu}}{|\vec{q}|^{2}}, \\
O^{(2)}_{\mu\nu} (q) &=& u_{\mu} u_{\nu}
-\frac{\vec{q}_{\mu}\vec{q}_{\nu}}{|\vec{q}|^{2}}
-\frac{q_{\mu}q_{\nu}}{q^{2}}, \\
O^{(3)}_{\mu\nu} (q) &=& \frac{q_{\mu}q_{\nu}}{q^{2}}, \\
O^{(4)}_{\mu \nu} (q) &=& O^{(2)}_{\mu \lambda} u^{\lambda}
\frac{q_{\nu}}{|\vec{q}|}
+\frac{q_{\mu}}{|\vec{q}|}u^{\lambda} O^{(2)}_{\lambda \nu},
\ea
are the three projectors of different types of gluon modes (``magnetic",
 ``electric", and unphysical ``longitudinal") \cite{us}, and one
intervening operator, respectively. By definition, $u_{\mu}=(1,0,0,0)$
and $\vec{q}_{\mu} = q_{\mu} -(u\cdot q) u_{\mu}$.

The current-current correlation function in Eq.~(\ref{cur-cur}) is
explicitly transverse. Similarly, the expression for the polarization
tensor that includes the contribution of the would be NG bosons is also
transverse. And, the corresponding gluon propagator reads
\ba
i{\cal D}^{(g)}_{\mu\nu} (q)&=& 
\frac{1}{q^{2}+\Pi_{1}}
O^{(1)}_{\mu\nu} 
+\frac{\Pi_{3}}{(q^{2}+\Pi_{2})\Pi_{3}+(\Pi_{4})^{2}}
O^{(2)}_{\mu\nu} \nonumber\\
&+&\frac{1}{\lambda q^{2}}
O^{(3)}_{\mu\nu}.
\label{gluon_propagator}
\ea
The poles of the correlation function (\ref{cur-cur}) and the gluon
propagator (\ref{gluon_propagator}) define the spectrum (as well as
screening effects) of the magnetic and electric type collective
excitations,
\ba 
q^{2} + \Pi_{1}(q) &=& 0, 
\quad \mbox{``magnetic"},
\label{spectrum-mag}\\
\left[q^{2} + \Pi_{2}(q)\right] \Pi_{3}(q) + [\Pi_{4}(q)]^{2} 
&=& 0, 
\quad \mbox{``electric"}. 
\label{spectrum-el}
\ea
To proceed with the analysis, we need to know the explicit expression for
the polarization tensor. The general finite temperature result of
Ref.~\cite{Rsch1} is rather complicated. We notice, however, that some
limiting cases are easily tractable \cite{GS-long}. Here we pay a special
attention to the nearcritical region ($T \alt T_{c}$ and $|\Delta_T|/T \ll
1$), where the CG modes are expected to appear.

Before presenting our main results, we would like to mention that the
properties of CG modes are extremely sensitive to the Landau damping
effects. A clear signal of the CG modes usually appears only in dirty
samples where Landau damping is rather inefficient \cite{CG}. In
the clean (no impurities) limit, on the other hand, Landau damping is
so strong that CG modes may become unobservable \cite{Tak2}.

Here we assume that quark matter in the CFL phase is a clean system.
This may or may not be true, because there could exist some natural
impurities in the vicinity of the critical point. However, if we show (as
we actually do) that CG modes exist in the clean limit where Landau
damping is strongest, then adding impurities into consideration can only
improve the quality of the CG modes.

In the nearcritical region, the calculation of the one-loop polarization
tensor [see Refs.~\cite{Rsch1,GS-long} for the general representation]
could be done approximately as an expansion in powers of $|\Delta_{T}|/T$.
The result reads
\ba
\Pi_{1} &\simeq & -\omega_{p}^{2} \left[
\frac{7\zeta(3)|\Delta_{T}|^{2}}{12\pi^{2}T^{2}}
+\frac{7}{3}v^{2}-\frac{7i\pi}{12}v \right. \nonumber \\
&+&\left. \frac{\pi v |\Delta_{T}| }{96 T} \left(5\pi +33 i
+10i\ln\frac{v}{2}\right) \right] ,
\label{Pi1-explicit}\\
\Pi_{2} &\simeq & \omega_{p}^{2} \left[ -3
+\frac{21\zeta(3)|\Delta_{T}|^{2}}{4\pi^{2}T^{2}}
+4v^{2}-\frac{7i\pi}{6}v \right. \nonumber \\
&+&\left. \frac{\pi v |\Delta_{T}| }{48 T} \left(9\pi +28 i
+18i\ln\frac{v}{2}\right) \right] ,
\label{Pi2-explicit}\\
\Pi_{3} &\simeq & \omega_{p}^{2} \left[
-\frac{7\zeta(3)|\Delta_{T}|^{2}}{12\pi^{2}T^{2}}
+\frac{2}{3}v^{2}+\frac{5i\pi|\Delta_{T}|}{48T}v \right] ,
\label{Pi3-explicit}\\
\Pi_{4} &\simeq & \frac{2}{3}\omega_{p}^{2} v,
\label{Pi4-explicit} 
\ea 
where $\omega_{p} =g_{s}\mu/\sqrt{2}\pi$ is the plasma frequency, $T$ is
the temperature, $|\Delta_{T}|$ is the value of the gap at this given
temperature, and $v\equiv q_{0}/|\vec{q}|$. In our derivation, we treated
both $|\Delta_{T}|/T$ and $v$ as small parameters of the same order. All
cubic and higher order terms has been dropped. We also restricted
ourselves to the low-energy region $|q_{0}|\ll |\Delta_{T}|$. Now, by
making use of the above explicit expressions for the polarization tensor
components, from Eq.~(\ref{spectrum-el}) we derive the approximate
dispersion relation of the electric type modes:
\be 
v_{cg}^2+i\frac{45 \pi|\Delta_{T}|}{224T} v_{cg}
-\frac{9\zeta(3)|\Delta_{T}|^2}{8\pi^2T^2} =0, 
\label{disp-eq} 
\ee
The solution to the dispersion equation (\ref{disp-eq}) reads 
\be 
q_{0} =\frac{|\Delta_{T}|}{T}(\pm x^{*} - i y^{*}) |\vec{q}|,
\label{result} 
\ee
where $x^{*} \approx 0.193$ and $y^{*}\approx 0.316$. This solution
corresponds to the gapless CG mode. The width of this mode is quite large,
but this is not surprising since the clean limit of quark matter is
considered. We have also derived an analytical expression which does not
assume the smallness of $v_{cg}$, and which includes all the corrections
up to order $(\Delta_{T}/T)^{2}$. Because of the complicated structure,
see Ref.~\cite{GS-long} for details, we do not present the corresponding
equation here. Instead, we describe the numerical solution in words. We
found that the ratio of the width, $\Gamma=-2 Im(v_{cg})|\vec{q}|$, to the
energy of the CG mode, $\varepsilon_{q}=Re(v_{cg})|\vec{q}|$, increases
when the temperature of the system goes away from the critical point. At
temperature $T^{*} \approx 0.986 T_{c}$, this ratio formally goes to
infinity. This value of $T^{*}$ could serve as an estimate of the
temperature at which the CG mode disappears. Of course, such an estimate
is rather crude because our approximations break before the temperature
$T^{*}$ is reached.

\begin{figure}
\epsfxsize=7.6cm 
\epsffile{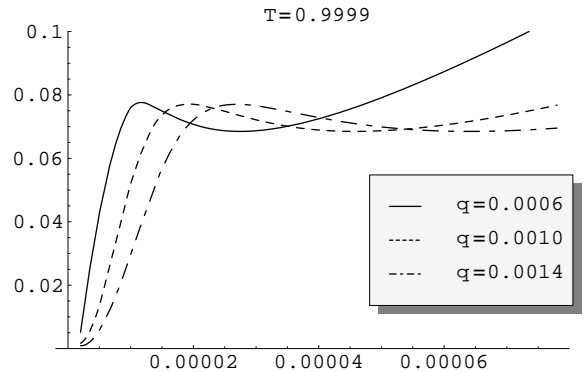} 
\caption{The spectral density of the electric modes at $T=0.9999 T_{c}$
(corresponding to $ |\Delta_{T}| \simeq 0.031 T_{c} $) as a function of
energy. Everything is measured in units of $T_{c}$.}
\label{fig:1}
\end{figure}

Because of the large imaginary part in Eq.~(\ref{result}), one
might conclude that the CG modes cannot not be observable in the
clean CFL phase of quark matter. By calculating the spectral
density of the electric type gluons in the close vicinity of the
critical temperature, we see that such a conclusion is not quite
correct. As is shown in Fig.~\ref{fig:1}, there is a well defined
peak in the spectral density which scales with the momentum in
accordance with a linear law. To plot the figure, we used the standard
Bardeen-Cooper-Schrieffer (BCS) dependence of the value of the gap on
temperature, obtained from the following implicit expression: 
\be
\ln\frac{\pi T_{c}}{e^{\gamma}|\Delta_{T}|}=\int_{0}^{\infty}
\frac{d\epsilon 
\left(1-\tanh\frac{1}{2T}\sqrt{\epsilon^2+|\Delta_{T}|^2}\right)}
{\sqrt{\epsilon^2+|\Delta_{T}|^2}}, 
\ee 
where $\gamma \approx 0.567$ is the Euler constant. As was shown in
Ref.~\cite{PR-weak}, such a dependence remains adequate in the case of a
color superconductor.
It has to be emphasized that the results in Eq.~(\ref{result}) and
Fig.\ref{fig:1} are obtained for a clean system where Landau damping
effects have their full strength. Therefore, it should have been expected
that the CG modes are overdamped. However, our analysis shows that the
presence of two different types of quarks with nonequal gaps in the CFL
phase plays an important role in partial suppression of Landau damping. We
checked that if all quarks had the same value of the gap in their spectrum
(as in ordinary metals), the ratio of the imaginary and real parts in the
solution [$v\simeq -14i\zeta(3)|\Delta_{T}|/3\pi^{3}T$ which is pure
imaginary in the approximation considered] would be much larger than 1,
meaning that the CG modes would be unobservable in clean systems.

Now, let us discuss the Meissner effect near $T_c$. This can be
done by examining the response of the quark system to an external
static magnetic field. So, we consider the magnetic $\Pi_{1}$
component of the polarization tensor in the limit $|q_{0}| \ll
|\vec{q}| \to 0$. By substituting $v=0$ on the right hand side of
Eq.~(\ref{Pi1-explicit}), we obtain
\be
\Pi_{1} \simeq
-\frac{7\zeta(3)|\Delta_{T}|^{2}}{12\pi^{2}T^{2}} \omega_{p}^{2}.
\label{static}
\ee
Since the right hand side is nonzero, it means that a static magnetic
field is expelled from the bulk of a superconductor. Thus, the
conventional Meissner effect is unaffected by the presence of the CG
modes. We would like to note that the scale, given by $\omega_{p}
|\Delta_{T}|/T$ in Eq.~(\ref{static}), is not directly related to the
penetration depth of the magnetic field. In order to determine the actual
depth, we would need to know the dependence of the $\Pi_{1}$ component for
a relatively wide range of spatial momenta, $0 < |\vec{q}| \alt \omega_{p}
|\Delta_{T}|/T$. Without doing the calculation here, we mention that the
penetration depth should be of order $\lambda_{P} \simeq
(|\Delta_{T}|\omega_{p}^{2})^{-1/3}$ (the Pippard penetration depth)
\cite{GS-long}.

In passing, we remind that, because of spontaneous breaking of chiral
symmetry, there are real NG bosons in the CFL phase of a color
superconductor. Unlike the scalar CG modes, the NG bosons are
pseudoscalars. According to Ref.~\cite{Zar}, their dispersion relation is
determined by the equation: $q^{\mu}\Pi_{\mu\nu}q^{\nu} \equiv
q^{2}\Pi_{3} =0$. In the nearcritical region, as follows from
Eq.~(\ref{Pi3-explicit}), this is equivalent to
\be
v_{ng}^{2}+i\frac{5\pi|\Delta_{T}|}{32T} v_{ng}
-\frac{7\zeta(3)|\Delta_{T}|^2}{8\pi^2T^2}=0, 
\label{disp-NG} 
\ee
which is qualitatively the same as Eq.~(\ref{disp-eq}), and it has a
solution with similar properties. In fact, the dispersion relation for the
NG bosons is given by an expression of the same type as that in
Eq.~(\ref{result}), but with $x^{*}\approx 0.215$ and $y^{*}\approx
0.245$. By comparing the spectrum of the CG modes with the spectrum of the
NG bosons, we see that they are quite similar. This observation may
provide some indirect justification to interpret the CG modes as
``revived"  NG bosons. Such an interpretation, however, should be used
with a caution.
\begin{figure}
\epsfxsize=7.6cm
\epsffile{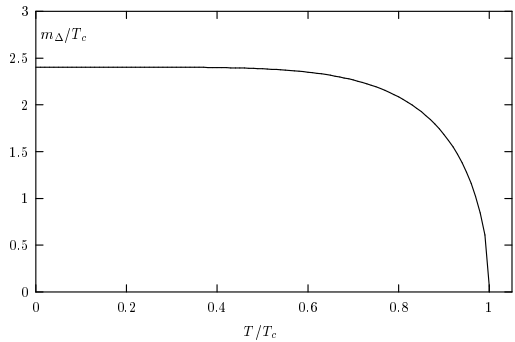}
\caption{Mass of the light massive gluon mode as a function of
temperature.}
\label{fig:2}
\end{figure}
For completeness, let us mention that there are two massive gluon modes in
the CFL phase \cite{GS-long}. One of them is a plasmon with the mass
$\omega_{p}$, and the other is a relatively light mode with the mass in
the range $|\Delta_{T}| < q_{0} < 2|\Delta_{T}|$. The plasmon has very
narrow width, while the other mode is stable. The dependence of the light
massive mode on temperature is graphically presented in Fig.~\ref{fig:2}.

In conclusion, by making use of an explicitly gauge covariant approach, we
predict the existence of the gapless CG modes in the CFL phase of cold
dense quark matter in the near-critical region (just below $T_{c}$) where
a considerable density of thermally excited quasiparticles is present. It
is important to mention that the presence of the CG modes coexists with
the usual Meissner effect, i.e., an external static magnetic field is
expelled from the bulk of a superconductor. The existence of the Meissner
effect is a clear signature that the system remains in the symmetry broken
(superconducting) phase.

In the case of the CFL phase, as we showed, the CG modes appear even in
the clean limit. Despite the sizable width, their traces can be observed
in the spectral density of the electric gluons.  Taking the effect of
impurities into account should, in general, make the CG modes more
pronounced \cite{Tak2}. In realistic systems such as compact stars,
natural impurities of different nature could further improve the
quality of the gapless CG modes.

The existence of a gapless scalar CG modes (in addition to the
pseudoscalar NG bosons) is a very important property of the color
superconducting phase. They may affect thermodynamical as well as
transport properties of the system in the nearcritical region.  In its
turn, this might have a profound effect on the evolution of forming
compact stars. The CG modes might also have a consequence on a possible
existence of the hypothetical quark-hadron continuity, suggested in
Ref.~\cite{SW-cont}. Indeed, one should notice that, in the hadron phase,
there does not seem to exist any low energy excitations with the quantum
numbers matching those of the CG modes.

In the future, it would be interesting to generalize our analysis to the
so-called $S2C$ phase of dense QCD. In absence of true NG bosons in the
$S2C$ phase, the gapless CG modes might play a more important role. Our
general observations suggest that Landau damping should have stronger
influence in the case of two flavors. At the same time, the presence of
massless quarks could lead to widening the range of temperatures where the
CG modes exist. To make a more specific prediction, one should study the
problem in detail.

{\bf Acknowledgments}.
Authors thank V.A.~Miransky and K.~Yamawaki for interesting discussions.
V.P.G. is grateful to S.~Sharapov for fruitful discussions on CG modes
in standard superconductors. I.A.S. is grateful to A.~Goldman for useful
comments.  This work is partially supported by Grant-in-Aid of Japan
Society for the Promotion of Science (JSPS) \#11695030. The work of V.P.G.
was also supported by the SCOPES projects 7~IP~062607 and 7UKPJ062150.00/1
of the Swiss NSF.  He wishes to acknowledge JSPS for financial support.
The work of I.A.S. was supported by the U.S. Department of Energy Grant
No.~DE-FG02-87ER40328.


\begin{references}


\item[$^{*}$]On leave of absence from Bogolyubov Institute for
Theoretical Physics, 252143, Kiev, Ukraine.

\bibitem{CG} R.V.~Carlson and A.M.~Goldman, \prl {\bf 34}, 11 (1975);
J. Low Tem. Phys. {\bf 25}, 67 (1976).

\bibitem{Kulik} I.O.~Kulik, O.~Entin-Wohlman, and R.~Orbach,
J. Low Tem. Phys. {\bf 43}, 591 (1981).

\bibitem{Tak1} K.Y.M.Wong and S.~Takada,
\prb {\bf 37}, 5644 (1988).

\bibitem{Tak2} Y.~Ohashi and S.~Takada,
J. Phys. Soc. Japan {\bf 66}, 2437 (1997);
\prb {\bf 62}, 5971 (2000).


\bibitem{SchSch} A.~Schmid and G.~Sch{\"o}n,
\prl {\bf 34}, 941 (1975).


\bibitem{BarFra} B.C.~Barrois, Nucl. Phys. {\bf B129}, 390 (1977).

\bibitem{Bail} D.~Bailin and A.~Love,
Phys. Rep. {\bf 107}, 325 (1984).

\bibitem{W1} M.~Alford, K.~Rajagopal, and F.~Wilczek,
\pl B{\bf 422}, 247 (1998).

\bibitem{S1} R.~Rapp, T.~Sch\"{a}fer, E.V.~Shuryak, and
M.~Velkovsky, \prl {\bf 81}, 53 (1998).

\bibitem{HIC-Stars}
M.~Alford, J.~A.~Bowers, and K.~Rajagopal,
J.\ Phys.\ {\bf G27}, 541 (2001).

\bibitem{ARW} M.~Alford, K.~Rajagopal, and F.~Wilczek,
Nucl. Phys. {\bf B537}, 443 (1999).

\bibitem{PR1} R.D.~Pisarski and D.H.~Rischke,
\prl {\bf 83}, 37 (1999).

\bibitem{us2} I.A.~Shovkovy and L.C.R.~Wijewardhana,
\pl B {\bf 470}, 189 (1999).

\bibitem{Spt} T.~Sch\"{a}fer,
Nucl. Phys. {\bf B575}, 269 (2000).

\bibitem{Son}
D.T.~Son, \prd {\bf 59}, 094019 (1999).

\bibitem{us} D.K.~Hong, V.A.~Miransky, I.A.~Shovkovy,
and L.C.R.~Wijewardhana, \prd {\bf 61}, 056001 (2000).

\bibitem{SW2} T.~Sch\"{a}fer and F.~Wilczek,
\prd {\bf 60}, 114033 (1999).

\bibitem{PR-weak}
R.~D.~Pisarski and D.~H.~Rischke,
Phys.\ Rev.\ D {\bf 61}, 074017 (2000).

\bibitem{Rsch1} D.H.~Rischke, \prd {\bf 62}, 054017 (2000).

\bibitem{Zar} K.~Zarembo, \prd {\bf 62}, 054003 (2000).

\bibitem{GS-long} V.P.~Gusynin and I.A.~Shovkovy,
preprint NUC-MINN-01/10-T.

\bibitem{bs-23} V.~A.~Miransky, I.~A.~Shovkovy, and L.~C.~Wijewardhana,
\prd {\bf 62}, 085025 (2000);
{\em ibid.} D {\bf 63}, 056005 (2001).

\bibitem{SW-cont}
T.~Sch\"{a}fer and F.~Wilczek,
Phys.\ Rev.\ Lett.\  {\bf 82}, 3956 (1999).


\end{references}
\end{document}